\documentclass[11pt,a4paper]{emulateapj}
\usepackage{verbatim}
\usepackage{epsfig}
\usepackage{amsmath}
\usepackage{natbib}

\begin{document}
 
\title{Simultaneous Linear and Circular Optical Polarimetry of Asteroid (4) Vesta}
\author{Sloane J. Wiktorowicz \& Larissa A. Nofi\altaffilmark{1}}
\affil{Department of Astronomy and Astrophysics, University of California, Santa Cruz, CA 95064}
\email{sloanew@ucolick.org}

\altaffiltext{1}{Present address: Institute for Astronomy, University of Hawaii, Honolulu, HI 96822}

\date{\today}

\begin{abstract}

From a single, 3.8-hour observation of asteroid (4) Vesta at $13\, \fdg7$ phase angle with the POLISH2 polarimeter at the Lick Observatory Shane 3-m telescope, we confirm rotational modulation of linear polarization in $B$ and $V$ bands. We measure the peak-to-peak modulation in degree of linear polarization to be $\Delta P = (294 \pm 35) \times 10^{-6}$ (ppm) and time-averaged $\Delta P / P = 0.0577 \pm 0.0069$. After rotating the plane of linear polarization to the scattering plane, asteroidal rotational modulation is detected with $12 \sigma$ confidence and observed solely in Stokes $Q/I$. POLISH2 simultaneously measures Stokes $I$, $Q$, $U$ (linear polarization), and $V$ (circular polarization), but we detect no significant circular polarization with a $1 \sigma$ upper limit of 140 ppm in $B$ band. Circular polarization is expected to arise from multiple scattering of sunlight by rough surfaces, and it has previously been detected in nearly all other classes of Solar System bodies save asteroids. Subsequent observations may be compared with surface albedo maps from the {\it Dawn Mission}, which may allow identification of compositional variation across the asteroidal surface. These results demonstrate the high accuracy achieved by POLISH2 at the Lick 3-m telescope, which is designed to directly detect scattered light from spatially unresolvable exoplanets. \\

\end{abstract}

\section{Introduction}

Optical polarimetry provides a unique view into the surface albedo, roughness, and composition of asteroids. The majority of polarimetric studies monitor polarization as a function of Sun-asteroid-observer phase angle $\alpha$. Due to the geometry of Main Belt asteroid orbits as seen from Earth, it is generally difficult to observe asteroids with $\alpha \ga 30^\circ$. For $\alpha \la 20^\circ$, coherent backscattering from rough regolith causes linear polarization parallel to the scattering plane (so-called ``negative branch polarization"), rather than perpendicular as for Rayleigh scattering (e.g., \citealp{Muinonen2015}). This effect is also observed in Mercury, the Moon, and Mars (e.g., \citealp{Lyot1929, Dollfus1974}). Fortuitously, the transition between negative and positive branch polarization may be observed from Earth for most asteroids, and the maximum absolute polarization in the negative branch $P_{\rm min}$, inversion phase angle $\alpha_{\rm inv}$, and polarimetric slope $h$ at the inversion phase angle allow differentiation between classes of asteroids (e.g., \citealp{Cellino2015}). \\

Less well-studied is rotational modulation in polarized light. While asteroid rotation may be readily apparent in photometry, polarimetric amplitudes are decreased and correspondingly difficult to measure. The highest significance detection of polarimetric rotational modulation is currently from asteroid (4) Vesta, due to its striking albedo heterogeneity \citep{Degewij1979, Broglia1989, Lupishko1999}. Since asteroidal polarization and albedo are generally anticorrelated (Umov effect), anticorrelation of simultaneous photometry and polarimetry suggests that surface roughness and composition are spatially homogeneous. However, departures from the Umov effect suggest that surface roughness and/or composition are heterogeneous. Thus, expansion of polarimetric rotational modulation detections will greatly improve understanding of asteroidal surface properties. For Vesta, the {\it Dawn Mission} has provided high spatial resolution maps \citep{Roatsch2013}, which allows unprecedented comparison of polarimetry with surface features. \\

Successful observations of polarimetric rotational modulation of Vesta have been performed at phase angles near $P_{\rm min}$, and rotation is observed to produce peak-to-peak $\Delta P / P \approx 0.08 - 0.14$ \citep{Degewij1979, Broglia1989}. Since $P_{\rm min} \approx -0.6\%$ \citep{Lupishko2014}, rotational modulation presents a peak-to-peak polarimetric modulation of $\Delta P \la 0.06\%$. As this is near the accuracy limit of conventional, waveplate polarimeters, other detections are few and at low SNR (e.g., (7) Iris: \citealp{Broglia1990}, (16) Psyche: \citealp{Broglia1992}, and (1943) Anteros: \citealp{Masiero2010}). \\

While optical circular polarization of Mercury, Venus, Earth's twilight sky, the Moon, Mars, Jupiter, Saturn, and comets has been detected, asteroids are conspicuously absent from the list \citep{Kemp1971a, Kemp1971b, Angel1972, Swedlund1972, Michalsky1974, Wolstencroft1976, Dollfus1987, Metz1987, Morozhenko1987, Rosenbush1997, Manset2000, Rosenbush2007}. Since asteroids tend to be fainter than Solar System planets, the combination of relatively large telescopes and highly accurate polarimeters is needed for detection of circular polarization. We discuss simultaneous linear and circular polarimetric observations of asteroid Vesta with the POLISH2 polarimeter at the Lick Observatory 3-m telescope on a single night. This instrument, primarily designed to search for scattered light from spatially unresolvable exoplanets, routinely reaches polarimetric sensitivity of order $10^{-6}$ (ppm) on naked-eye stars. \\

\section{Methods}
\subsection{The POLISH2 Polarimeter}

POLISH2 (POlarimeter at Lick for Inclination Studies of Hot jupiters 2) is a dual photoelastic modulator (PEM) polarimeter mounted at straight Cassegrain focus of the Lick Observatory Shane 3-m telescope. Downstream of the telescope secondary, two Hinds Instruments fused silica PEMs (I/FS40 and I/FS50) are located to minimize instrumental polarization. The compression/extension axis of the first PEM is oriented parallel to Celestial North ($+Q$), while the second PEM is oriented Northwest ($-U$). Acoustic resonance in the PEMs is induced, and their stress birefringence simultaneously modulates incident linear and circular polarization. Therefore, each Stokes parameter ($Q$, $U$, and $V$) is modulated at specific frequencies.  \\


Downstream of the PEMs is a filter wheel with Bessell $UBV$ filters. Instrumental polarization due to the filters is discussed later in this section. Next, a two-wedge Wollaston prism bifurcates the beam into opposite linear polarization states, where polarization incident on the PEMs is AC coupled into an intensity modulation. Both modeling and telescope flat-field lamp observations show that polarization is modulated at sum and difference linear combinations of the PEM fundamental frequencies ($f_1 = 40$ kHz, $f_2 = 50$ kHz). Indeed, Stokes $Q$ is modulated primarily at 100 kHz, $U$ at 10 and 90 kHz, and $V$ at 30, 50, and 130 kHz. \\

Both Wollaston output beams are focused at $15\arcsec$ diameter field stops and field lenses, and light is detected by Hamamatsu H10721-110 SEL photomultiplier tubes (PMTs). A pair of Stanford Research SR570 current preamplifiers converts the current output of the PMTs into a voltage that is digitized by a National Instruments PXIe-6124 analog to digital converter. Square wave voltage outputs from both PEM controllers are simultaneously digitized by this device, and the fundamental frequencies and phases of both PEMs are precisely measured in each 5 second data file. This allows frequent, precise knowledge of the modulation frequencies. \\

Since POLISH2 is an AC-coupled polarimeter, a lock-in amplifier approach is taken in software \citep{Hough2006, Wiktorowicz2008}. Briefly, both a sine and cosine are constructed for each modulation frequency, and the product of these sinusoids with mean-subtracted, raw data from each PMT are calculated over 0.1 second data segments. The time average of each product is then taken, which removes the contribution of raw data varying at frequencies other than the modulation frequency. These time-averaged values, the so-called ``in-phase" ($X_i$) and ``quadrature" ($Y_i$) components of the power modulated at each frequency $i$, are the observable quantities: \\

\begin{eqnarray}
X_i = 2 \langle \sin(2 \pi f_i t) [ S(t) - DC ] \rangle, \\
Y_i = 2 \langle \cos(2 \pi f_i t) [ S(t) - DC ] \rangle.
\end{eqnarray}

\noindent Here, $S(t)$ and $DC = \langle S(t) \rangle$ are raw and time-averaged voltages from each PMT. \\

 \citet{Wiktorowicz2008} show that polarized intensity (Stokes $Q$, $U$, or $V$) scales with $R_i = \sqrt{X_i^2 +Y_i^2}$, while total intensity (Stokes $I$) scales with $DC$. Since fractional polarization is the quantity of astrophysical interest, we define normalized Stokes parameters $q \equiv Q/I$, $u \equiv U/I$, and $v \equiv V/I$ for clarity. Each of these is composed of $x_i \equiv X_i / DC$ and $y_i \equiv Y_i / DC$, which are measured at the appropriate modulation frequency $i$. Thus, fractional polarizations are given by the following: \\

\begin{eqnarray}
q_0 & = & \pm \sqrt{x_q^2 + y_q^2}, \label{q0} \\
u_0 & = & \pm \sqrt{x_u^2 + y_u^2}, \\
v_0 & = & \pm \sqrt{x_v^2 + y_v^2}, \\
p_0 & = & \sqrt{q_0^2 + u_0^2}, \label{p0}
\end{eqnarray}

\noindent where equation \ref{p0} describes the degree of linear polarization. \\

\begin{deluxetable*}{c c c c c c c}
\tabletypesize{\normalsize}
\tablecaption{Telescope Polarization Observations}
\tablewidth{0pt}
\tablehead{
\colhead{Dataset} & \colhead{HR} & \colhead{$q$} & \colhead{$u$} & \colhead{$p$} & \colhead{$\Theta$} & \colhead{$v$} \\
& & (ppm) & (ppm) & (ppm) & ($^\circ$) & (ppm)}
\startdata
Star + TP	& 1791		& +56.4(5.0)	& +67.3(4.3)		& 87.7(4.6)	& 25.0(1.5)	& $-73$(11) \\
$\cdots$	& 4295		& +57.3(7.0)	& +52.8(6.1)		& 77.7(6.6)	& 21.3(2.4)	& $-$155(16) \\
$\cdots$	& 4357		& +63.8(8.1)	& +56.3(7.0)		& 84.8(7.6)	& 20.7(2.5)	& $-$68(18) \\
$\cdots$	& 4534		& +66.2(6.7)	& +73.3(5.8)		& 98.6(6.2)	& 23.9(1.8)	& $-$136(15) \\
$\cdots$	& 4540		& +94(16)		& +91(14)			& 130(15)		& 21.9(3.3)	& $-$82(36) \\
$\cdots$	& 5235		& +63(10)		& +59.7(9.0)		& 86.1(9.8)	& 21.8(3.2)	& $-$63(23) \\
$\cdots$	& 5435		& +56(10)		& +50.9(9.0)		& 75.2(9.8)	& 21.1(3.7)	& $-$86(23) \\
\hline
Mean TP	&  			& +61.0(2.9)	& +63.5(2.5)		& 88.0(2.7)	& 23.06(88)	& $-$99.0(6.4) \\
\hline
Star		& 1791		& $-$4.6(5.8)	& +3.8(5.0)	& 4.3(5.5)		& 70(34)		& +26(13) \\
$\cdots$	& 4295		& $-$3.7(7.6)	& $-$10.7(6.6)	& 9.1(6.7)		& 125(22)		& $-$56(17) \\
$\cdots$	& 4357		& +2.8(8.6)	& $-$7.1(7.4)	& 5.1(7.6)		& 146(45)		& +31(19) \\
$\cdots$	& 4534		& +5.2(7.3)	& +9.8(6.3)	& 9.0(6.5)		& 31(21)		& $-$37(16) \\
$\cdots$	& 4540		& +33(17)		& +27(14)		& 40(16)		& 20(11)		& +17(37) \\
$\cdots$	& 5235		& +2(11)		& $-$3.8(9.3)	& 2.3(9.6)		& $-$		& +36(24) \\
$\cdots$	& 5435		& $-$5(11)	& $-$12.6(9.3)	& 10.1(9.5)	& 125(28)		& +13(24)
\enddata
\label{tp}
\end{deluxetable*}

It has long been known that the magnitude of a vector is biased toward larger values due to the uncertainty in vector components, especially for low SNR measurements. Thus, the subscript 0 in Equations \ref{q0} through \ref{p0} indicates the naive, biased estimator. To minimize this bias, we use the generalized MAS estimator  \citep{Plaszczynski2014}: \\

\begin{eqnarray}
q & = & \pm \left[ q_0 - b^2 \left(\frac{1 - \exp (-q_0^2 / b^2)}{2 q_0} \right) \right], \label{unbias1} \\
b^2 & = &  \frac{x^2 \sigma_y^2 + y^2 \sigma_x^2}{x^2 + y^2}, \label{unbias2} \\
\sigma_q^2 & = & \frac{x^2 \sigma_x^2 + y^2 \sigma_y^2}{x^2 + y^2}. \label{unbias3} 
\end{eqnarray}

\noindent Here, $q$ is an estimate of the true, normalized Stokes parameter and $\sigma_q^2$ is an estimate of its variance. Similar equations exist for Stokes $u$ and $v$. The position angle of linear polarization is then given by $\Theta = \onehalf \arctan(u / q)$. \\

To calibrate instrumental efficiency factors in each $UBV$ filter, a linear polarizer and quarter-wave Fresnel rhomb are used to polarize and inject lamp light into POLISH2 with $\sim 100\%$ Stokes $q$, $u$, and $v$. Instrumental polarization, caused by the filter between the PEMs and Wollaston, is assessed by rotating the filter wheel to a clear aperture. With $\sim 100\%$ $q$ (or $u$) injected, removal of the $B$ filter causes a difference of $< 200$ ppm in $u$ (or $q$). Therefore, we expect astrophysical sources, where linear polarization is a maximum of order $1\%$, to experience a negligible $\sim 1$ ppm crosstalk between $q$ and $u$. \\

\subsection{Observations}

For sky subtraction, the telescope is nodded by $30\arcsec$ between target and sky fields, where two, 30-second target integrations are taken for each 30-second sky field integration to maximize SNR. We observed asteroid Vesta for 3.82 hours on the night of 2014 March 19 UT, which was composed of 14 minute integrations per band in a repeated, $BVU$ filter sequence. Since POLISH2 is an aperture-integrated polarimeter with a $15\arcsec$ diameter field of view, the telescope tracked Vesta using JPL HORIZONS non-sidereal rates. For Vesta, sky-subtracted values of $x_i$ and $y_i$ are binned during each 14-minute integration. Non-zero telescope polarization, caused by reflectivity variations across the telescope mirrors, is subtracted via measurements of seven bright, unpolarized stellar calibrators. Each calibrator star is observed for 9 minutes. These stars are identified both from \citet{Bailey2010} and our unpublished survey of bright stars with POLISH2 on the Lick Observatory 1-m telescope. \\

\begin{deluxetable*}{c c c c c c c}
\tabletypesize{\normalsize}
\tablecaption{$UBV$ Telescope Polarization}
\tablewidth{0pt}
\tablehead{
\colhead{UT Date} & \colhead{Band} & \colhead{$q$} & \colhead{$u$} & \colhead{$p$} & \colhead{$\Theta$} & \colhead{$v$} \\
(2014) & & (ppm) & (ppm) & (ppm) & ($^\circ$) & (ppm)}
\startdata
April 20, 21	& $U$	& +128(19)	& +101(16)		& 162(18)		& 19.2(3.0)	& $-$125(45) \\
March 19		& $B$	& +61.0(2.9)	& +63.5(2.5)		& 88.0(2.7)	& 23.06(88)	& $-$99.0(6.4) \\
April 20, 21	& $B$	& +51.8(5.7)	& +60.5(4.9)		& 79.5(5.2)	& 24.7(1.9)	& $-$4(12) \\
April 20, 21	& $V$	& +7.4(7.3)	& +48.7(6.9)		& 48.7(6.9)	& 40.7(4.3)	& $-$80(17) \\
\enddata
\label{tpvar}
\end{deluxetable*}

Instrumental $+Q$ is nominally aligned with Celestial North. Comparison of the position angles of five strongly polarized stars with the catalogs of \citet{Heiles2000} ($\epsilon$ Aur, $\zeta$ Oph, 53 Per, 55 Cyg, and HD 154445), observed during two runs between 2014 March 19 and 2014 April 21 UT, indicates $\Theta_{\rm POLISH2} - \Theta_{\rm catalogs} = -1\, \fdg08 \pm 0\, \fdg 34$. Thus, instrumental $+Q$ differs from catalogs at the $1^\circ$ level with $3.2\sigma$ confidence, indicating the absolute accuracy in position angle is $\sim 1^\circ$\label{rotoff}. The positioning accuracy of the Lick 3-m Cassegrain rotator is limited to $\sim 0\, \fdg3$, which may be the cause of the slight offset. \\

Fortuitously, Vesta was observed in the range $\Phi = 90\, \fdg 571 \pm 0\, \fdg 067$, where $\Phi$ is the angle between the Vesta-Celestial North and Vesta-Sun directions. Thus, rotation of observed Stokes $q$ and $u$ (in the celestial frame) to the scattering plane (denoted with primes, $q'$ and $u'$) involves a rotation of only $| \Theta - \Theta' | \approx 0\, \fdg5$, which is accomplished by well-known relations (e.g., \citealp{Bagnulo2006}). \\

\section{Results}
\subsection{Telescope Polarization}
\label{telescopepoln}

We measure $B$ band telescope polarization (TP) from observations of seven unpolarized stars taken during the same night as the Vesta observations (Table \ref{tp}). Quantities in parenthesis indicate $1 \sigma$ uncertainties in the last two digits. Telescope polarization is calculated from the weighted mean of $x$ and $y$ observables across all stars for each Stokes parameter. We detect significant telescope linear (and circular) polarization with $28 \sigma$ ($12 \sigma$) confidence and 3 ppm (7 ppm) accuracy, illustrating the simultaneous linear and circular polarimetric sensitivity of POLISH2. \\

Table \ref{tp} also lists telescope polarization-subtracted, $B$ band polarization of these unpolarized calibrator stars. For no star is linear polarization detected with $\geq 3 \sigma$ confidence, which suggests all stars to be viable calibrators for telescope polarization. However, circular polarization of the peculiar A star HR 4295 is detected with $3.3 \sigma$ confidence, and its lack of significant linear polarization suggests circular polarization to be intrinsic to the system as opposed to linear-to-circular conversion by the ISM \citep{Martin1972}. As it is a marginal detection, we do not attempt to further explain this result. \\

Telescope polarization was only measured in $B$ band on the night of Vesta observations, but Vesta data were taken in $UBV$ bands. To measure and correct for the wavelength dependence of telescope polarization, we utilize $UBV$ measurements of HR 4295, 4534, 6212, and 6220 obtained on 2014 April 20 and 21 UT (Table \ref{tpvar}). Since $q$ and $u$ vary by $< 3 \sigma$ between March and April, we use the difference between April $UBV$ measurements to estimate $U$ and $V$ band telescope polarization in March. Inexplicably, $B$ band circular polarization measurements differ significantly between these runs, but as section \ref{circpol} shows, this discrepancy lies within the uncertainty in Vesta's circular polarization. \\

\begin{figure}
\centering
\includegraphics[scale=0.7]{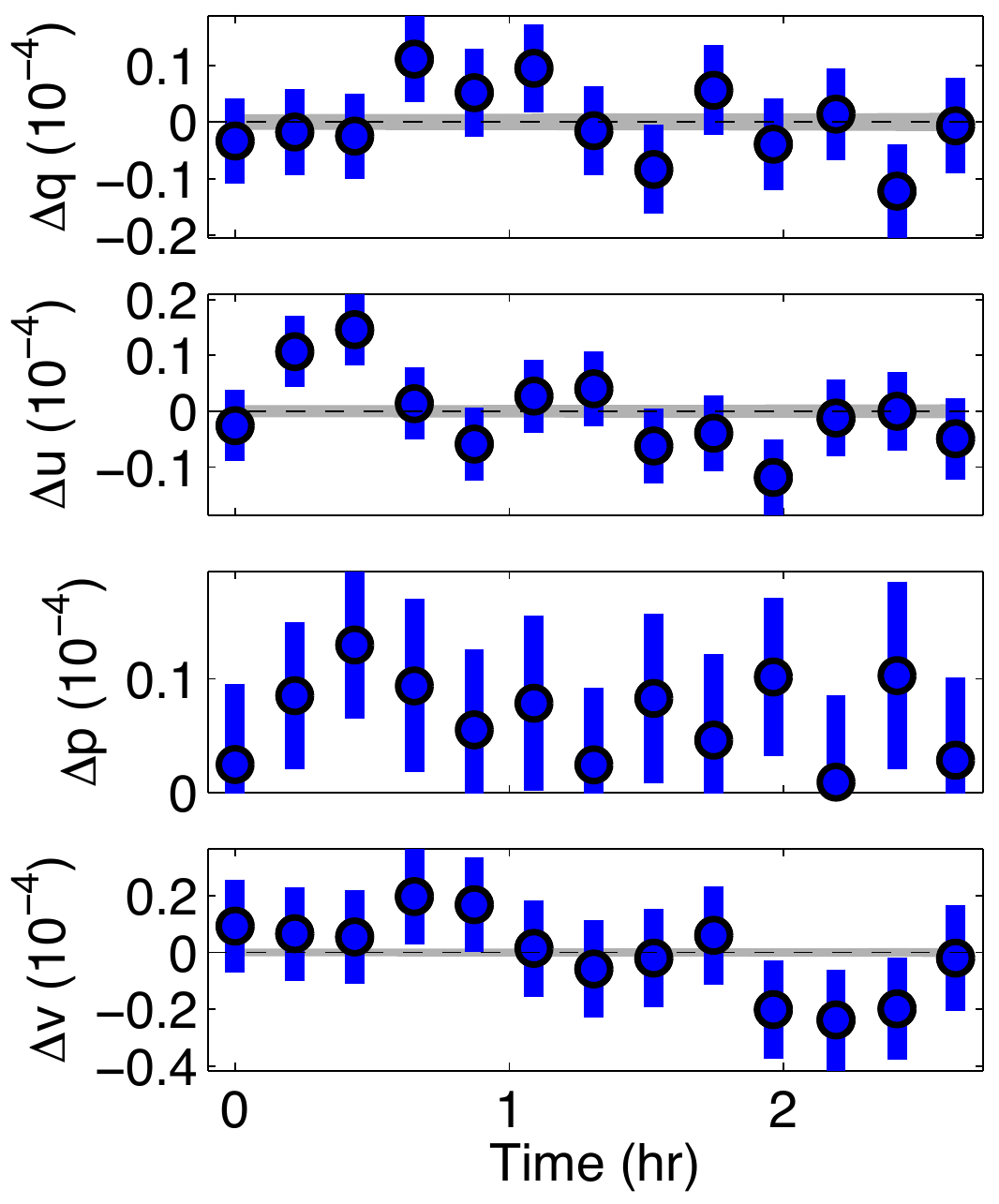}
\caption{$B$ band observations of HR 4295 in 13-minute bins. Grey bands indicate the photon noise level, which is $\sim 2$ ppm in each Stokes parameter.}
\label{stability}
\end{figure}

\begin{figure*}
\centering
\includegraphics[scale=0.7]{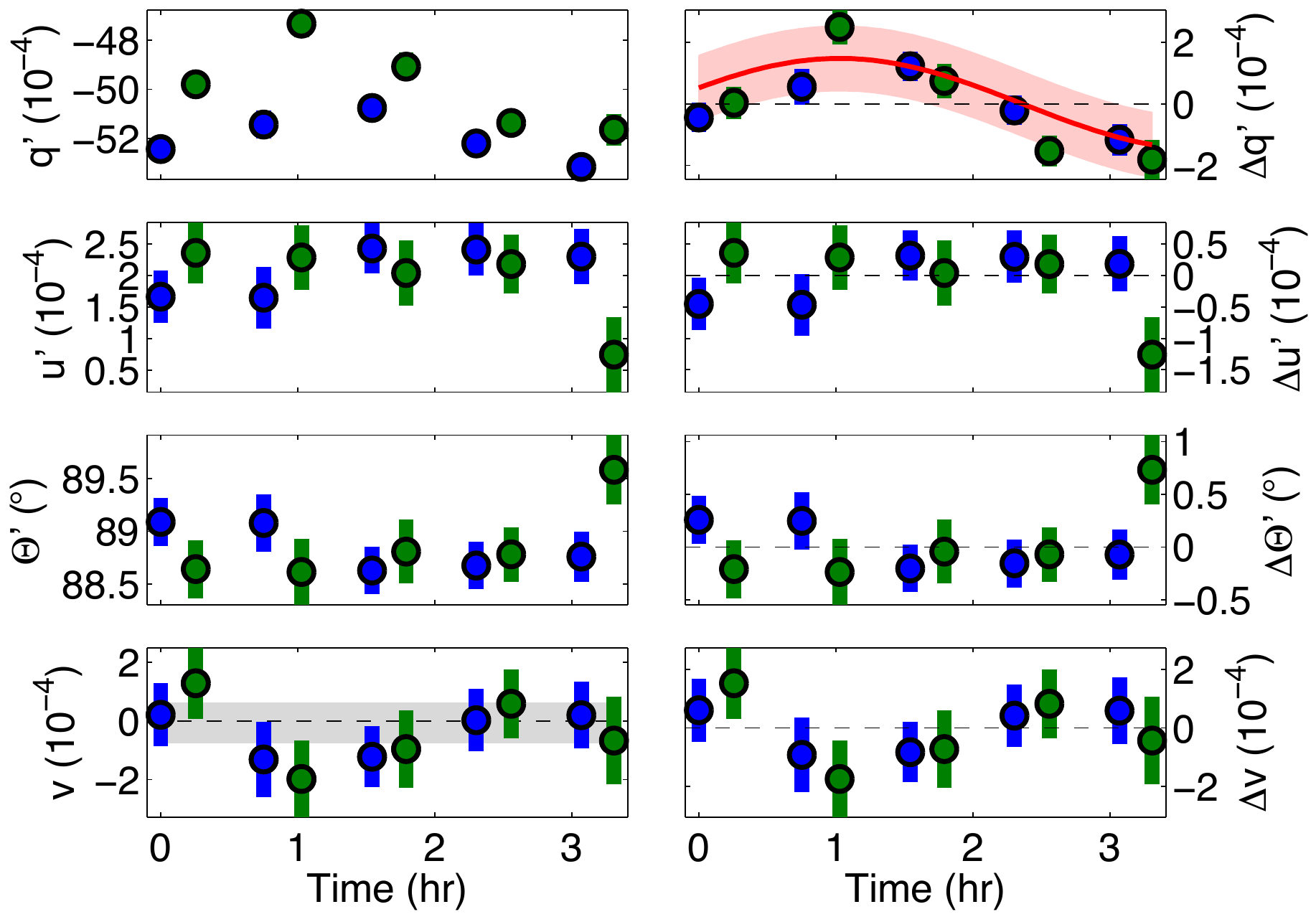}
\caption{{\it Left}: Telescope polarization-subtracted, Lick 3-m Vesta observations in $B$ (blue) and $V$ bands (green). The grey region in the Stokes $v$ panel indicates the $\pm 1\sigma$ range for unpolarized calibrator stars (Table \ref{tp}). {\it Right:} Data subtracted by the time average in each band. Best-fit sinusoids are tied to Vesta's known, 5.342-hour rotation period. Bin duration is 14 minutes with median uncertainty 52, 47, 120 ppm, and $0\, \fdg27$ in Stokes $q'$, $u'$, $v$, and $\Theta'$, respectively. Median photon noise is $\sim 20$ ppm per bin across all Stokes parameters. Pink shading ($\Delta q'$ panel) indicates $\pm 1 \sigma$ uncertainty in sinusoidal fitting to combined $BV$ band data (red curve).}
\label{vesta}
\end{figure*}

\subsection{Instrumental Stability}

Figure \ref{stability} shows a continuous, 2.85-hour observation of the unpolarized calibrator star HR 4295 taken in $B$ band on 2014 March 12 UT. Measurement sensitivity, the ability to measure change, is given by median $q$, $u$, and $v$ uncertainties in 13-minute bins (7.8, 6.7, and 17 ppm, respectively). Measurement accuracy, the ability to reject non-astrophysical change, is given by the square root of the weighted variance from bin to bin (6.4, 7.0, and 13 ppm in Stokes $q$, $u$, and $v$, respectively). Since measurement sensitivity and accuracy are essentially identical, a constant nature may only be rejected with $0.3 \sigma$, $1.1 \sigma$, and $0.2 \sigma$ confidence from a $\chi^2$ test. Therefore, over the timescale of Vesta observations, we demonstrate that POLISH2 is stable to 7.0 ppm for linear polarization and to 13 ppm for circular polarization. \\

\begin{deluxetable*}{c c c c c c c}
\tabletypesize{\normalsize}
\tablecaption{Asteroid (4) Vesta Observations}
\tablewidth{0pt}
\tablehead{
\colhead{UT Time} & \colhead{Band} & \colhead{$q$} & \colhead{$u$} & \colhead{$p$} & \colhead{$\Theta$} & \colhead{$v$} \\
(2014 March 19) & & (ppm) & (ppm) & (ppm) & ($^\circ$) & (ppm)}
\startdata
10:10	& $U$		& $-6,870(300)$	& +550(300)	& 6,880(300)	& 87.7(1.2)	& $-760(780)$ \\
10:57	& $\cdots$	& $-6,360(230)$	& +80(210)	& 6,360(230)	& 89.62(95)	& $-760(560)$ \\
11:42	& $\cdots$	& $-6,880(230)$	& +550(210)	& 6,900(230)	& 87.70(89)	& $+180(570)$ \\
12:28	& $\cdots$	& $-6,950(260)$	& +390(250)	& 6,950(260)	& 88.4(1.0)	& $-880(650)$ \\
13:14	& $\cdots$	& $-6,150(280)$	& +1,000(260)	& 6,220(280)	& 85.4(1.2)	& $-1,080(690)$ \\
\hline
09:39	& $B$		& $-5,243(48)$		& +167(42)	& 5,246(48)	& 89.09(23)	& $+20(110)$ \\
10:24	& $\cdots$	& $-5,144(58)$		& +165(49)	& 5,146(58)	& 89.08(27)	& $-130(130)$ \\
11:12	& $\cdots$	& $-5,077(47)$		& +243(40)	& 5,082(47)	& 88.63(23)	& $-120(100)$ \\
11:57	& $\cdots$	& $-5,221(48)$		& +241(41)	& 5,226(48)	& 88.68(23)	& $+0(110)$ \\
12:43	& $\cdots$	& $-5,316(51)$		& +230(44)	& 5,321(51)	& 88.76(24)	& $+20(110)$ \\
\hline
09:55	& $V$		& $-4,980(52)$		& +236(48)	& 4,985(52)	& 88.64(28)	& $+130(120)$ \\
10:41	& $\cdots$	& $-4,733(56)$		& +229(51)	& 4,738(56)	& 88.61(31)	& $-200(130)$ \\
11:27	& $\cdots$	& $-4,908(56)$		& +204(52)	& 4,912(56)	& 88.81(30)	& $-100(130)$ \\
12:13	& $\cdots$	& $-5,137(50)$		& +219(46)	& 5,141(50)	& 88.78(26)	& $+60(120)$ \\
12:58	& $\cdots$	& $-5,165(64)$		& +75(60)		& 5,165(64)	& 89.58(33)	& $-70(150)$ \\
\enddata
\label{vestatab}
\end{deluxetable*}

\subsection{Vesta Linear Polarization}

Table \ref{vestatab} and Figure \ref{vesta} (left column) detail our telescope polarization-subtracted, simultaneous Stokes $q'$, $u'$, and $v$ measurements of Vesta in $B$ and $V$ bands. Note that circular polarization $v$ is invariant under rotation to the scattering plane. We confirm that Vesta's time-averaged linear polarization is parallel to the scattering plane ($\Theta' \approx 90^\circ$), which is consistent with the phase angle of the observations. We discover that the magnitude of Vesta's time-averaged degree of negative-branch polarization decreases with increasing wavelength with $14 \sigma$ confidence ($-6,622 \pm 110$, $-5,180 \pm 22$, and $-4,992 \pm 25$ ppm in $UBV$ bands, respectively). The wavelength dependence of asteroidal linear polarization depends on taxonomic class and phase angle of observation (e.g., \citealp{Cellino2015}). \\

Figure \ref{vesta} (right column) shows mean-subtracted polarization in each band to illustrate the amplitude of rotational modulation, which is immediately apparent in Stokes $q'$ ($\Delta q'$ panel). Here, a constant nature is rejected with $4.7 \sigma$ confidence from a $\chi^2$ test of $q_x$ and $q_y$ data ($\chi^2 = 55.6$, $\nu = 16$). We therefore fit sinusoids with Vesta's rotational period of 5.342 hours to our mean-subtracted $BV$ band data (due to faintness in $U$ band), which provides a detection of rotational modulation with $12 \sigma$ confidence. We find a peak-to-peak modulation of $\Delta p = 294 \pm 35$ ppm and a time-averaged polarization degree of $\langle p \rangle = -5,091 \pm 17$ ppm. Therefore, the ratio of peak-to-peak modulation to the time average is $\Delta p / \langle p \rangle = 0.0577 \pm 0.0069$. \\

After rotation to the scattering plane, Stokes $u'$ is expected to be zero for single scattering events. Figure \ref{vesta} ($u'$ panel) shows that the weighted mean position angle of Vesta observations ($\Theta'$ panel) differs from $90^\circ$ by $-1\, \fdg161 \pm 0\, \fdg082$. This is consistent with the magnitude and sense of the $-1\, \fdg08 \pm 0\, \fdg 34$ rotational offset of POLISH2 in the plane of the sky, which is likely due to inaccuracy in the telescope Cassegrain rotator (Section \ref{rotoff}). \\

\subsection{Vesta Circular Polarization}\label{circpol}

Since time-averaged $\langle v \rangle = -105 \pm 37$ ppm, we detect no significant broadband, circular polarization of Vesta and set a $1\sigma$ upper limit of $|v| < 140$ ppm. In addition, we detect no significant rotational modulation of circular polarization. \\

\section{Discussion}

Previous measurements suggest peak-to-peak $\Delta p / \langle p \rangle = 0.08-0.24$ \citep{Degewij1979, Broglia1989, Lupishko1999}. Since our measurements of $\Delta p / \langle p \rangle$ differ with only $1.7 \sigma$ confidence between $B$ and $V$ bands, we combine $B$ and $V$ band data to find $\Delta p / \langle p \rangle = 0.0577 \pm 0.0069$, at the minimum of the reported range. We observe the bright, nearly unpolarized star HR 4295 to be non-variable to 7.0 ppm and 13 ppm, in linear and circular polarization, respectively, over a similar timescale to the Vesta observations. This demonstrates the extreme stability of POLISH2 for time-domain science. \\

Significant Stokes $u'$ is detected even after rotation to the scattering plane, but we attribute this to the $\sim 1^\circ$ rotational offset between the orientations of instrumental and celestial $+q$. This is likely due to the known inaccuracy in rotational positioning of the Lick 3-m telescope. \citet{Lupishko1999} observe temporal variations in polarization position angle $\Theta'$ whose amplitude decreases with increasing wavelength ($\Delta \Theta' = 8^\circ$ to $2\, \fdg5$ from $U$ to $I$ bands, respectively), which is attributed to surface features. However, a $\chi^2$ analysis shows that our $B$ and $V$ band position angles are inconsistent with a constant with only $0.7 \sigma$ and $1.3 \sigma$ confidence, respectively. The peak-to-peak modulations of sinusoidal fits to $\Theta'$ are $0\, \fdg46$ and $0\, \fdg77$, respectively; therefore, we observe no significant rotation of polarization position angle above $\sim 1^\circ$. \\

This may be due to the phase angle of observation, as \citet{Lupishko1999} observe closer to the phase angle where $P = P_{\rm min}$, and they report the extremely large value of $\Delta p / \langle p \rangle = 0.24$. Additionally, our observations only sample 72\% of Vesta's rotational period, so spatial features contributing to the \citet{Lupishko1999} result may not have been visible during our study. Thus, repeated observations near $P_{\rm min}$, and sampling the full rotation of the asteroid, are necessary. While POLISH2 is simultaneously sensitive to linear and circular polarization, we find no evidence of significant circular polarization with $|v| < 140$ ppm ($1 \sigma$). \\

Since POLISH2 is primarily designed to directly detect polarized, scattered light from spatially unresolvable exoplanets, the accuracy required for this task lies at the part-per-million level. With $\sim 4$ hours of observations of the $V = 6.2$ mag object Vesta, we demonstrate an uncertainty in time-averaged linear polarization of 17 and 15 ppm in Stokes $q$ and $u$, respectively. Therefore, POLISH2 at the Lick 3-m telescope is well-suited to the study of the brightest exoplanets in scattered light. \\

\section{Conclusion}
\label{section_conclusion}

We confirm rotational modulation of asteroid (4) Vesta's optical linear polarization in $B$ and $V$ bands with $12 \sigma$ confidence, and we measure a peak-to-peak modulation of $\Delta p = 294 \pm 35$ ppm. This is in stark contrast to observations of a nearly unpolarized calibrator star, which are constant to 7.0 ppm and 13 ppm in linear and circular polarization over a similar timescale as the Vesta observations. We observe modulation relative to the time-averaged degree of linear polarization to be $\Delta p / \langle p \rangle = 0.0577 \pm 0.0069$, which is lower than previously reported and may be due to the phase angle of observation. Our observations were taken with the asteroid on the negative polarization branch, and they are consistent with linear polarization parallel to the scattering plane. We discover that the time-averaged degree of linear polarization differs between $U$ and $V$ bands with $14 \sigma$ confidence, where the magnitude of negative-branch polarization decreases with wavelength. \\

We detect no significant rotation of polarization position angle with time to $\sim 1^\circ$, inconsistent with \citet{Lupishko1999}, but this may be due to the difference in phase angle between our studies and also to our incomplete coverage of the asteroid's rotation. We detect no significant circular polarization of Vesta with a $1 \sigma$ upper limit of $|v| < 140$ ppm. \\

Given that the apparent magnitude of Vesta is similar to the brightest known host stars of spatially unresolvable exoplanets, we demonstrate the extreme accuracy of POLISH2 in the time domain for direct, scattered-light detection of such exoplanets. Indeed, the peak-to-peak modulation of Vesta's linear polarimetric modulation is measured with 35 ppm accuracy over $\sim 4$ hours, which suggests that repeated observations of bright exoplanet host stars reaches the accuracy necessary to enable exoplanet detections. \\

\acknowledgments
The authors would like to thank the referee, Stefano Bagnulo, as well as Alberto Cellino, Daniel Jontof-Hutter, Ludmilla Kolokolova, and Joseph Masiero for valuable discussions. We would also like to acknowledge the tireless efforts of the Lick Observatory staff. This work was performed (in part) under contract with the California Institute of Technology (Caltech) funded by NASA through the Sagan Fellowship Program executed by the NASA Exoplanet Science Institute. We acknowledge support from the NASA Origins of Solar Systems program through grant NNX13AF63G.  \\
 
 {\it Facilities:} \facility{Shane (POLISH2)}.


 \end{document}